\documentclass{ifacconf}
\usepackage{graphicx}      
\usepackage{svg}
\usepackage{natbib}        
\usepackage{array}
\usepackage{booktabs}
\usepackage{tcolorbox}
\usepackage{amssymb}
\usepackage{balance}
\usepackage{amsmath,amssymb,amsfonts}
\usepackage{algorithmic}
\usepackage{textcomp}
\usepackage[dvipsnames]{xcolor}
\usepackage{csquotes}
\def\BibTeX{{\rm B\kern-.05em{\sc i\kern-.025em b}\kern-.08em
    T\kern-.1667em\lower.7ex\hbox{E}\kern-.125emX}}
\usepackage{mathtools}

\usepackage{circuitikz}
\begin{document}
\begin{frontmatter}

\title{Identification of Port-Hamiltonian Differential-Algebraic Equations from Input-Output Data\thanksref{footnoteinfo}} 

\thanks[footnoteinfo]{This work is part of the DAMOCLES research project which received funding from the Eindhoven Artificial Intelligence Systems Institute, as part of the EMDAIR funding program. Furthermore, this work is funded by the European Union (ERC, COMPLETE, 101075836). Views and opinions expressed are however those of the author(s) only and do not necessarily reflect those of the European Union or the European Research Council Executive Agency. Neither the European Union nor the granting authority can be held responsible for them.}

\author[First]{N. Hagelaars} 
\author[First]{G.J.E. van Otterdijk} 
\author[First]{S. Moradi}
\author[First,Third]{R. T\'{o}th}
\author[Fourth]{N.O. Jaensson}
\author[First]{M. Schoukens}

\address[First]{Control Systems Group, Eindhoven University of Technology, Eindhoven, the Netherlands.}
\address[Third]{Systems and Control Laboratory, Institute for Computer Science and Control, Budapest, Hungary.}
\address[Fourth]{Processing and Performance of Materials Group, Eindhoven University of Technology, Eindhoven, the Netherlands.}

\begin{abstract}                
Many models of physical systems, such as mechanical and electrical networks, exhibit algebraic constraints that arise from subsystem interconnections and underlying physical laws. Such systems are commonly formulated as differential-algebraic equations (DAEs), which describe both the dynamic evolution of system states and the algebraic relations that must hold among them. Within this class, port-Hamiltonian differential-algebraic equations (pH-DAEs) offer a structured, energy-based representation that preserves interconnection and passivity properties. This work introduces a data-driven identification method that combines port-Hamiltonian neural networks (pHNNs) with a differential-algebraic solver to model such constrained systems directly from noisy input–output data. The approach preserves the passivity and interconnection structure of port-Hamiltonian systems while employing a backward Euler discretization with Newton’s method to solve the coupled differential and algebraic equations consistently. The performance of the proposed approach is demonstrated on a DC power network, where the identified model accurately captures system behaviour and maintains errors proportional to the noise amplitude, while providing reliable parameter estimates. 
\end{abstract}

\begin{keyword}
System Identification, port-Hamiltonian models, Differential Algebraic Equations. 
\end{keyword}

\end{frontmatter}

\section{Introduction}
Modelling dynamical systems based on known first-principle relations and data is essential for understanding and controlling complex systems across scientific and engineering disciplines. These tasks, however, are complicated by nonlinearity, system complexity, and sensitivity to noise \citep{SchoukensLjung2019}. The choice of modelling approach plays a significant role in addressing these challenges in the modelling process, exploiting measured data, which is also known as system identification.

A popular choice is grey-box modelling, which integrates first-principles-based knowledge with empirical data. These methods essentially combine the interpretability of white‐box modelling with the adaptability of black‐box modelling, reducing data requirements by incorporating known physical structure. Recent efforts have focused on embedding physics-based insights into data-driven methods, aiming to fuse adaptability with physical interpretability.

The port-Hamiltonian framework serves as a grey-box modelling method by embedding known physical laws into a structured model in terms of energy, interconnection and passivity \citep{vanderSchaft2014Port-hamiltonianOverview}. This structure ensures consistency with first-principles modelling while allowing data-driven parameter estimation. To exploit the strength of port-Hamiltonian representations, in machine learning, Hamiltonian neural networks (HNNs) \citep{Greydanus-HNN} have been introduced, integrating the energy-based structure into learning algorithms to improve interpretability and robustness of the resulting models.

However, HNNs do not account for dissipation or external inputs, limiting their applicability to energy-conservative systems. Extensions to include dissipation \citep{sosanya2022dissipativehamiltonianneuralnetworks} and external inputs \citep{Zhong2020Symplectic} led to the development of port-Hamiltonian neural networks (pHNNs) \citep{Desai2021-PHNN-explicit-time-dependent}, and further, to composite pHNNs \citep{neary-compositePHNN-ODE}, which parametrise dissipation, inputs, and the Hamiltonian functions directly. These model structures, however, do not incorporate measurement noise, nor can be estimated with the existing methods using input-output (IO) data. To address this, output-error pHNNs (OE-pHNNs) were proposed \citep{moradi2025porthamiltonianneuralnetworksoutput}, enabling robust modelling with noisy measured IO data through an output-error structure.

Most of the aforementioned research focuses on ordinary differential equations (ODEs), but many real-world systems involve algebraic constraints due to interconnected subsystems or conservation laws, which require a differential-algebraic equation (DAE) formulation \citep{Mehrmann2023DifferentialalgebraicStructure}.
Moreover, in large‑scale networks, it is rarely sufficient to identify a single global ODE model. It is more common to focus on smaller subsystems to limit data requirements and maintain computational feasibility \citep{composite-PH-ID,HaberVerhaegen}. Modelling these subsystems accurately becomes challenging when algebraic constraints are present. Working directly with DAEs not only enforces these constraints, but also preserves the original sparsity and interconnection structure. This is a significant advantage for large‑scale engineering systems such as power grids or robotics. Consequently, extending port-Hamiltonian systems to DAEs yields port-Hamiltonian differential-algebraic equations (pH-DAEs), preserving the energy-based structure in the presence of constraints.

Several recent studies address data-driven identification of DAE systems. For instance, \cite{koch2025learningneuraldifferentialalgebraic} integrate DAEs by extending neural timesteppers, while \cite{Moya2023DAE-PINN:Networks} propose DAE physics-informed neural networks (PINNs) to solve differential and algebraic components simultaneously, avoiding the common practice of sequential solving. However, these methods do not capture the port-
Hamiltonian structure. In the pH-DAE context, \cite{zaspel2024} used Gaussian processes to extend pH-ODE models to nonlinear pH-DAEs, focusing on identifying subsystems from input and state-space data. \cite{neary2025neuralporthamiltoniandifferentialalgebraic} proposed neural pH-DAEs, but used index reduction to convert DAEs to ODEs for training. This index‑reduction step can be convenient, but it can introduce hidden constraints and compromise both numerical stability and parameter identifiability in the subsystems to be identified.

Building on these developments, this paper aims to further bridge the gap between physics-based and data-driven methodologies by focusing on the identification of pH-DAE systems from input-output data. The main contributions of this paper are:
\begin{itemize}
\item \textbf{Simultaneous solution of differential and algebraic equations:} a data-driven port-Hamiltonian formulation is proposed that solves both parts of the DAE system implicitly and consistently, without requiring index reduction.
\item \textbf{Preservation of physical structure:} the proposed formulation maintains the energy-based, port-Hamiltonian representation, ensuring interpretability, passivity, and correct interconnection structure throughout identification.
\item \textbf{Robust identification from noisy input–output data:} the framework directly models measurement noise within an output-error setting, enabling reliable identification of dynamical systems with algebraic constraints using only input–output measurements, without needing state measurements.
\end{itemize}

 The paper is structured as follows: Section \ref{sec:PHS} covers the port-Hamiltonian theory and the chosen representation. Section \ref{sec:identification} proposes an identification method based on a proposed pH-DAE model structure and an appropriate numerical solver. Section \ref{sec:sim-study} demonstrates applicability of the proposed method via a simulation study. Finally, Section \ref{sec:conclusion} summarises the main findings.

\section{Linear Time-Invariant Port-Hamiltonian Systems} \label{sec:PHS}
Port-Hamiltonian systems theory fundamentally relies on network modelling to represent complex physical systems. 
At their core, port-Hamiltonian systems integrate the classical Hamiltonian formulation of dynamics with the notion of power ports, 
thereby providing a structured representation of energy exchange and interconnection among subsystems \citep{vanderSchaft2013Port-HamiltonianSystems}. 
This energy-based representation ensures that interconnections preserve power and that dissipation is explicitly modelled, 
leading to physically interpretable and passive system descriptions.

A general representation of a linear time-invariant (LTI) pH-DAE is given by
\begin{subequations}
    \begin{align}
        E\dot{x}(t) &= (J - R)Qx(t) + Gu(t), \\
        y (t)&= G^\top Qx(t),
    \end{align}
    \label{eq:LTI-pHDAE}
\end{subequations}
where $x(t) \in \mathbb{R}^n$ is the state vector, and $u(t), y(t) \in \mathbb{R}^m$ are the input and output vectors, respectively, and $t\in\mathbb{R}$ is the time; in port-Hamiltonian systems, each input–output pair forms a power port, so $\dim u = \dim y = m$.
The matrices $E, Q, J, R \in \mathbb{R}^{n \times n}$ and $G \in \mathbb{R}^{n \times m}$ denote the \textit{descriptor}, \textit{energy}, \textit{structure}, \textit{dissipation}, and \textit{input} matrices, respectively, and satisfy
\[
J = -J^\top, \qquad R = R^\top \succeq 0, \qquad E^\top Q = Q^\top E \succeq 0.
\]
These structural properties ensure that interconnections preserve power and that the system remains passive. 

When $E$ is singular, the system simultaneously contains differential and algebraic equations, thereby constituting a DAE. The matrix $J$ encodes the interconnection topology and conserves energy through its skew-symmetry, while $R$ captures dissipation. The matrix $Q$ defines the energy variables and scales the Hamiltonian, and $G$ represents the external power port through which the system interacts with its environment.

The total stored energy of the system is described by the \textit{quadratic Hamiltonian}
\begin{equation}
    H(x) = \tfrac{1}{2}x^\top Q^\top E x,
\end{equation}
which satisfies $H(x) \ge 0$ when $E^\top Q = Q^\top E \succeq 0$. When $Q = I$, it follows that $E = E^\top \succeq 0$ \citep{Beattie2018LinearSystems}.

\section{Identification Approach} \label{sec:identification}
Prior identification methods for linear port-Hamiltonian systems either rely on frequency-response measurements \citep{benner2020identification} or reconstruct state trajectories from power-balance relations under access to resistive variables \citep{LTV}. These approaches, however, address pH-ODE models and do not extend to descriptor formulations with algebraic constraints. 

This section presents the proposed identification approach for pH-DAE systems. The considered system class and parametrised model structure are introduced in Sections~\ref{subsec:system_class} and~\ref{subsec:model_structure}, respectively. The identification algorithm based on the subspace encoder and output-error formulation is described in Section~\ref{subsec:id_algorithm}, followed by the solver algorithm employing a backward Euler discretization with Newton’s method for consistent integration of differential and algebraic equations in Section~\ref{subsec:solver}.
\subsection{Data-generating system}\label{subsec:system_class}
The considered systems, which obey the linear port-Hamiltonian dynamics with algebraic constraints \eqref{eq:LTI-pHDAE}, can be represented in terms of continuous-time state-space equations as
\begin{subequations}
    \begin{align}
        E\dot{x}(t) &= (J-R)Qx(t)+Gu(t), \\ 
        y(t) &= G^\top Qx(t), \\ 
        y_k &=y(kT_s)+v_k,
    \end{align} 
\end{subequations}
 where, the measured outputs $y_k$ are sampled with sampling rate $T_s$ and are assumed to contain measurement noise, $v_k$, a zero-mean i.i.d. noise with finite variance.

The collected sampled input-output measurements are defined as
\begin{equation*}
    \mathcal{D}_N = \{(u(kT_s),y(kT_s)\}^{N-1}_{k=0}.
\end{equation*}

\subsection{Model structure} \label{subsec:model_structure}
In order to model the dynamics of the considered systems, an output-error model structure is adopted. The model has the same pH-DAE form as~\eqref{eq:LTI-pHDAE},  with system matrices replaced by parametrised matrices $(E_\theta, J_\theta, R_\theta, Q_\theta, G_\theta);
\theta \in \Theta \subseteq \mathbb{R}^{n_\theta}.$ The parametrisation preserves the port-Hamiltonian structure for all
$\theta$ by enforcing
\[
J_\theta = \tfrac12(M_{J,\theta} - M_{J,\theta}^\top), \quad
R_\theta = L_{R,\theta} L_{R,\theta}^\top, \quad
E_\theta = L_{E,\theta} L_{E,\theta}^\top,
\]
where \(M_{J,\theta}\), \(L_{R,\theta}\), and \(L_{E,\theta}\) are
independently parametrised matrices. Throughout this work, we assume \(Q_\theta = I\).

\subsection{Identification algorithm} \label{subsec:id_algorithm}
Based on the parametrised port-Hamiltonian model introduced in Section~\ref{subsec:model_structure}, the unknown parameters $\theta$ are estimated by minimising the squared error between the measured $y_k$ and simulated output $\hat{y}_k$ of the model over the dataset $\mathcal{D}_N$:
\begin{equation}
    V_{\mathcal{D}_N}(\theta) = \frac{1}{N}\sum_{k=1}^{N-1}\|y_k - \hat{y}_k\|_2^2.
\end{equation}
We consider only measurement noise and therefore adopt an output-error formulation. Approaches that explicitly model process disturbances in DAE systems~\citep{process_noise, Stochastic_process_noise} follow a different stochastic framework and do not apply to the setting considered here.
To efficiently estimate parameters with the output measurement noise, the subspace encoder-based identification method (SUBNET)~\citep{SUBNET} is employed. This approach, extended to pHNNs in~\citep{moradi2025porthamiltonianneuralnetworksoutput}, divides the dataset into multiple subsections of truncation length $T$, allowing the averaged prediction loss (in this case simulation loss) to be minimised more efficiently:
\begin{subequations}
\label{eq:lossfunction}
\begin{align}
    V^{\mathrm{sub}}_{\mathcal{D}_N}(\theta,\eta)
    &= \frac{1}{C}\sum_{\tau=n+1}^{N-T+1}\sum_{k=0}^{T-1}
       \!\left\|y_{\tau+k}-\hat{y}_{\tau+k|\tau}\right\|_2^2,
       \label{eq:loss_a}
\end{align}
\vspace{1ex}
\noindent{subject to:}
\vspace{-1ex}
\begin{align}
    \hat{x}_{\tau|\tau} &= \psi_\eta(u_{\tau-n}^{\tau-1},y_{\tau-n}^{\tau}), 
       \label{eq:loss_b}\\
    E_\theta\hat{x}_{\tau+k+1|\tau} &=
       \mathrm{DAEsolve}\!\left([J_\theta{-}R_\theta]Q_\theta \hat{x}_{\tau+k|\tau}
       + G_\theta u_{\tau+k}\right), 
       \label{eq:loss_c}\\
    \hat{y}_{\tau+k|\tau} &= G_\theta^\top Q_\theta\hat{x}_{\tau+k|\tau}.
       \label{eq:loss_d}
\end{align}
\end{subequations}
In this notation, the time indices $k$ and $\tau$ refer to samples on the uniform grid $t = \ell T_s$, $\ell \in \mathbb{N}$ (e.g.\ $x_k = x(kT_s)$), where $\tau$ marks the beginning of a subsection. The term $\hat{x}_{\tau+k|\tau}$ denotes the predicted state at time $t = (\tau+k)T_s$, calculated by propagating the model with the DAE solver (Section~\ref{subsec:solver}) from the initial state at $t = \tau T_s$.

To estimate the initial state within a subsection containing $T$ samples, starting at $\tau \in [n,\, N{-}T]$ (denoted as $\hat{x}_{\tau|\tau}$), we introduce the parameterised encoder function $\psi_\eta(\cdot)$, which infers $\hat{x}_{\tau|\tau}$ from the past $n$ input–output samples. Here, $u_{\tau-n}^{\tau-1}=[u_{\tau-n}^\top\ \cdots\ u_{\tau-1}^\top]^\top$ 
with $y_{\tau-n}^{\tau}$ defined in the same way. 

In practice, the full loss~\eqref{eq:lossfunction} is approximated by randomly sampling a subset of subsections (mini-batches):
\begin{equation}
\label{eq:batch_loss}
V^{(\mathrm{sub,batch})}_{\mathcal{D}_N}(\theta,\eta)
= \frac{1}{N_{\mathrm{batch}}}\sum_{\tau\in\mathcal{I}}\frac{1}{T}
\sum_{k=0}^{T-1}\!\left\|y_{\tau+k}-\hat{y}_{\tau+k|\tau}\right\|_2^2,
\end{equation}
where $\mathcal{I}\subset\{n{+}1,\ldots,N{-}T{+}1\}$ is the set of sampled subsections and $N_{\mathrm{batch}}=|\mathcal{I}|$.  
The batch formulation uses the same model constraints as in~\eqref{eq:lossfunction}, 
while enabling efficient stochastic gradient optimisation using Adam optimizer~\citep{ADAM}.

\begin{figure*}[t!]
    \centering
    \includegraphics[width=0.75\textwidth, clip]{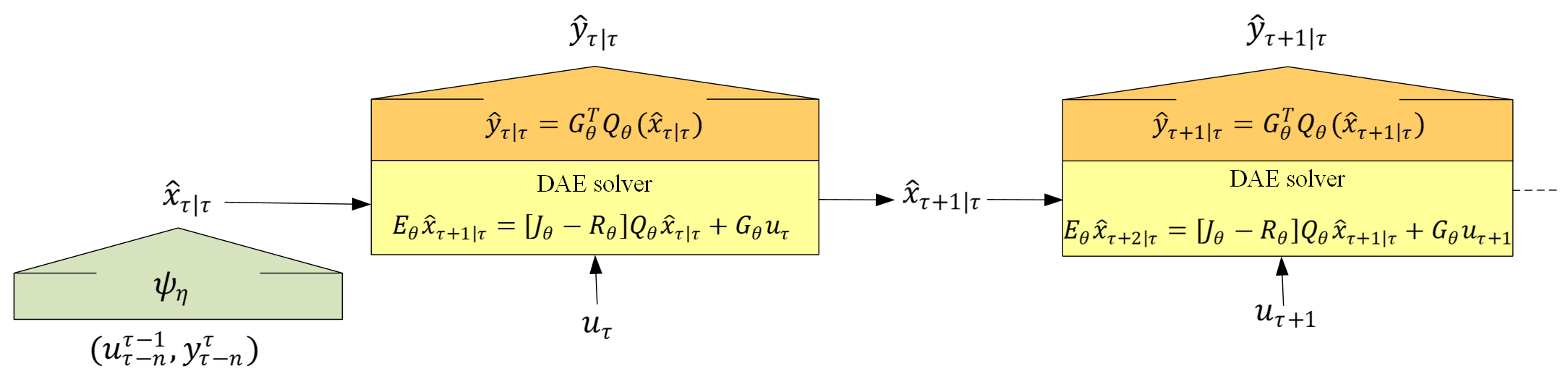}
    \caption{Computational pipeline of the simulated response using SUBNET and the DAE solver. The encoder $\psi_\eta$ estimates the initial state from past inputs and outputs, which is propagated by the DAE solver over the truncation length~$T$.}
    \label{fig:daesolver-schematic}
\end{figure*}

The computation of the simulated response consists of three steps:
\begin{enumerate}
    \item the encoder estimates the initial state for each subsection;
    \item the DAE solver estimates the state trajectory using the parametrised dynamics; and
    \item the output equation maps the simulated states to the corresponding outputs.
\end{enumerate}
The parameters $(\theta,\eta)$ are jointly optimised by minimising the loss~\eqref{eq:batch_loss} over selected subsections. 
Figure~\ref{fig:daesolver-schematic} illustrates the identification process, where the encoder $\psi_\eta$ estimates the initial state and the DAE solver advances the system dynamics over each truncation horizon~$T$.

\subsection{Computation of the model response} \label{subsec:solver}
Within the identification approach of Section~\ref{subsec:id_algorithm}, the state trajectory is propagated using the parametrised port-Hamiltonian model introduced in Section \ref{subsec:model_structure}.
To integrate both differential and algebraic components, an implicit DAE solver based on the backward Euler method is applied to the model.
The backward Euler method is employed as a first-order implicit scheme, offering stable integration for index-1 systems introduced in Section \ref{subsec:system_class}.

For a fixed sampling period $h = T_s$, let $\hat{x}_n \equiv \hat{x}_{\tau+k+1|\tau}$ with $t_n = t_{k+1}$. 
Applying the backward Euler method to the parametrised model
 yields
\begin{subequations}
\label{eq:BE-solver}
\begin{align}
E_\theta \tfrac{\hat{x}_n - \hat{x}_{n-1}}{h}
- (J_\theta - R_\theta) Q_\theta \hat{x}_n
- G_\theta u_n &= 0,\label{subeq:1_BE}\\ 
\hat{y}_n &= G_\theta^\top Q_\theta \hat{x}_n.
\label{subeq:2_BE}
\end{align}
\end{subequations}

The algebraic equation~\eqref{subeq:1_BE} is rewritten in terms of the residual vector $r(\hat{x}_n) \in \mathbb{R}^{n_x}$,
\begin{equation}
\label{eq:residual}
r(\hat{x}_n)
= \Bigl[
\tfrac{E_\theta}{h}\hat{x}_n
- \tfrac{E_\theta}{h}\hat{x}_{n-1}
- (J_\theta - R_\theta)Q_\theta \hat{x}_n
- G_\theta u_n
\Bigr],
\end{equation}
and the equation $r(\hat{x}_n) = 0$ is solved at each step using Newton’s method, initialised by the encoder estimate $\psi_\eta(\cdot)$ from Section~\ref{subsec:id_algorithm}. 
The residual equations enforce both differential and algebraic constraints on the state variables.
At each Newton iteration:
\begin{subequations}\label{eq:newton}
\begin{align}
J_r(\hat{x}_i)\,\Delta \hat{x}_i &= -\,r(\hat{x}_i),\\
\hat{x}_{i+1} &= \hat{x}_i + \Delta \hat{x}_i,
\end{align}
\end{subequations}
where $J_r(\hat{x}_i)=\partial r(\hat{x}_i)/\partial \hat{x}_i$, and $\Delta\hat{x}_i$ is the update step.
The iteration proceeds until $\|\hat{x}_{i+1}-\hat{x}_i\|_2 < \varepsilon$, in which $\varepsilon$ is the error tolerance.

Specifying the residual vector for the backward Euler equation~\eqref{subeq:1_BE} gives
\begin{subequations}
\label{eq:jacobian}
\begin{align}
    r(\hat{x}_n) &= J_r \hat{x}_n + b,\\
    J_r &= \tfrac{1}{h}E_\theta - J_\theta Q_\theta + R_\theta Q_\theta,\\
    b &= -\tfrac{1}{h}E_\theta \hat{x}_{n-1} - G_\theta u_n.
\end{align}
\end{subequations}

All solver operations, including residual evaluation and Jacobian computation, are implemented as differentiable mappings, enabling gradient propagation through the solver during optimisation of the loss~\eqref{eq:batch_loss}. 
Model and encoder parameters $(\theta, \eta)$ are updated jointly using stochastic gradient descent, e.g., the Adam method~\citep{ADAM}.
\paragraph*{Remark on Consistency.}
The backward Euler–Newton scheme is a convergent, A-stable implicit integrator for index-1 DAEs \citep{ascher1998computer}. Since it provides a convergent numerical approximation of the underlying pH-DAE dynamics, the consistency proof established for the OE-pHNN approach \citep{moradi2025porthamiltonianneuralnetworksoutput} remains applicable in the present setting.

\section{Simulation Study}\label{sec:sim-study}
To demonstrate the performance of the proposed identification approach, a DC power network is considered. The example, adapted from \citep{Mehrmann2019Structure-preservingSystems}.

\begin{figure}[b!]
    \centering
    \resizebox{\linewidth}{!}{%
        \begin{circuitikz}[american, >=latex]
                \draw
                (0,0) node[ground]{} to[V=$E_G$, invert] (0,2)
                to[R=$R_G$, i>^=$I_G$] (2,2) node[above]{$V_1$}
                to[L=$L$]                (4,2)
                to[R=$R_L$, -, i>^=$I$]  (6,2) node[above]{$V_2$}
                to[short]                          (8,2)
                to[R=$R_R$, i<^=$I_R$]             (8,0) node[ground]{};
            \draw (2,2) to[C=$C_1$, i<^=$I_1$] (2,0) node[ground]{};
            \draw (6,2) to[C=$C_2$, i<^=$I_2$] (6,0) node[ground]{};
        \end{circuitikz}
    }
  \caption{DC power network example as proposed by \cite{Mehrmann2019Structure-preservingSystems}.}
  \label{fig:PN_example}
\end{figure}
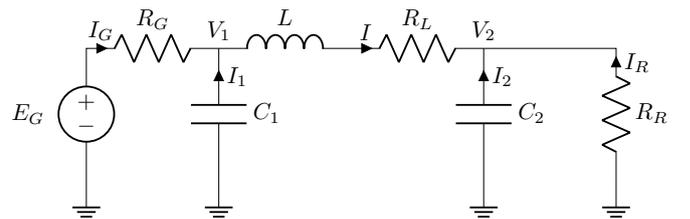

\subsection{System description}\label{subsec:data_generation}
The circuit consists of a DC generator with voltage source $E_G$, connected to a load through a $\pi$-model transmission line (Fig. \ref{fig:PN_example}). With $R_G,R_L,R_R>0$, $L>0$, and $C_1,C_2>0$, the system dynamics follow from Kirchhoff’s laws:
\begin{subequations}
\label{eq:PN-example-DAE}
    \begin{align}
        L\dot{I}&=-R_LI+V_2-V_1, \\
        C_1\dot{V}_1&=I-I_G, \\
        C_2\dot{V}_2&=-I-I_R, \\
        0&=-R_GI_G+V_1+E_G,\\
        0&=-R_RI_R+V_2.
    \end{align}
\end{subequations}
Setting $x=(I,V_1,V_2,I_G,I_R)$, $E=\text{diag}(L,C_1,C_2,0,0)$, $R=\text{diag}(R_L,0,0,R_G,R_R)$, $G=(0,0,0,1,0)^\top$, $Q=I$, $u=E_G$,
\begin{equation*}
    J = \begin{bmatrix}
        0 & -1 &1 &0 &0 \\
        1 &0 &0 &-1 &0 \\
        -1 &0 &0 &0 &-1 \\
        0 &1 &0 &0 &0 \\
        0 &0 &1 &0 &0 \\
    \end{bmatrix}, 
\end{equation*}
and $y=I_G$, the system is written as a linear descriptor pH-DAE, introduced in Eq.(\ref{eq:LTI-pHDAE}). The energy of the system is stored in the inductor and the capacitors and described by the quadratic Hamiltonian,
\begin{equation}
    \mathcal{H}(I,V_1,V_2)= \frac{1}{2}LI^2 + \frac{1}{2}C_1V_1^2 + \frac{1}{2}C_2V_2^2.
\end{equation}

For identification, the interconnection and input matrices $(J,Q,G)$ are fixed by the known topology, whereas the unknown component values are collected in $\theta = \{L,\,C_1,\,C_2,\,R_L,\,R_G,\,R_R\}$.
Only $E_\theta$ and $R_\theta$ are optimised, preserving the physical structure and ensuring interpretable parameter estimation.

\subsection{Dataset and model training}
To evaluate the proposed identification approach, the system in Fig.~\ref{fig:PN_example} is simulated using the nominal component values 
$\theta^*=\{L:2.0,\,C_1:0.01,\,C_2:0.02,\,R_L:1.0,\,R_G:6.0,\,R_R:3.0\}$. 
The objective is to generate input–output data representative of realistic operating conditions.

A multisine voltage excitation is applied as the generator input,
\begin{equation}
    u(t) = \sum^{40}_{i=1} \sin(2\pi if_0t+\phi_i),
\end{equation}
with $f_0=0.1$ and random phases uniformly sampled from $[0,2\pi)$. The system is simulated for $50$ seconds, with a sample rate $T_s=0.005$s, resulting in a dataset of $10000$ samples. Gaussian white noise, with a signal-to-noise ratio (SNR) of 20dB, is added to the sampled output signal to simulate measurement noise. 
The resulting dataset, containing only input–output measurements (i.e., no state observations), is used for identification.

\begin{figure}
    \centering
    \includegraphics[width=\linewidth]{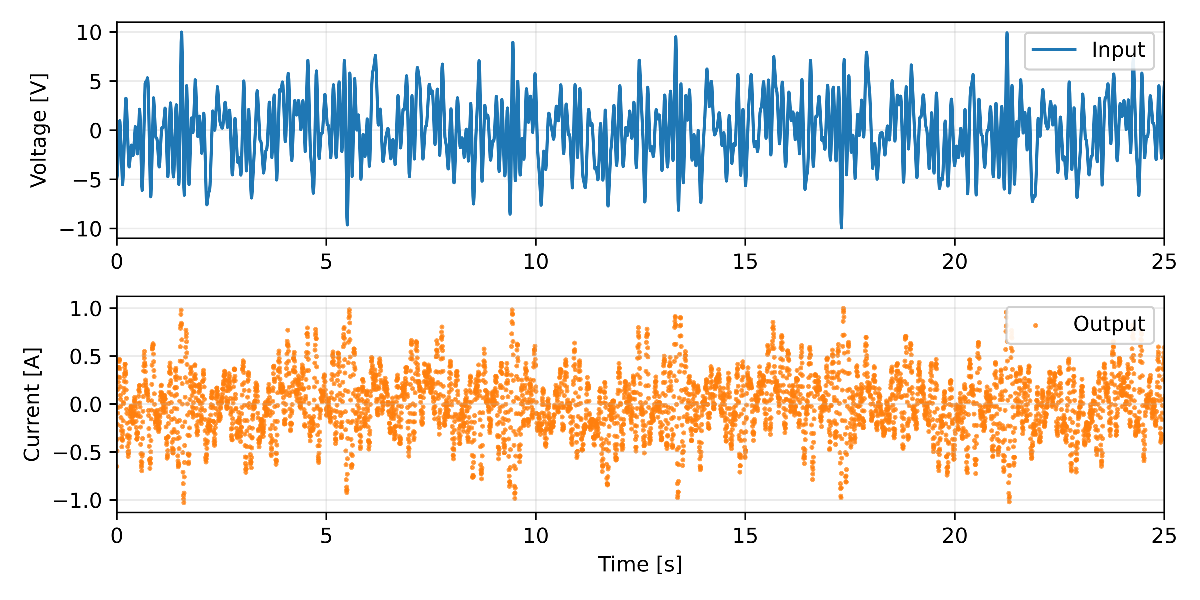}
    \caption{Example of the power network system behaviour for the first 25 seconds of the training dataset. The first subplot shows the external generator voltage taken as an input, $u=E_G$. The second subplot shows the corresponding noisy output measurement of the generator current, $y=I_G$.}
    \label{fig:inout}
\end{figure}

Three datasets were generated with different initial conditions of the system and different realisations of the input signal. One dataset was used for training the model, one set for validation, and one set for testing. Fig. \ref{fig:inout} shows the input and noisy output of the power network system.

Model training follows the procedure in Section~\ref{sec:identification}, using a linear encoder and the ADAM optimiser. 
Training is performed in batches of 256 samples with a learning rate decaying from $10^{-2}$ to $10^{-3}$ over 600–1000 epochs. 
The best model is selected based on the validation loss. 
Model performance is quantified by the Normalised Root Mean Square error (NRMS),
\begin{equation}
    \mathrm{NRMS} = \frac{\mathrm{RMS}}{\sigma_y}
    = \frac{\sqrt{\frac{1}{N}\sum_{k=1}^N \|y_k - \hat{y}_k\|_2^2}}{\sigma_y}.
\end{equation}
where $\hat y_k$ is the simulated output, $y_k$ the measured output at given time-samples, and $\sigma_y$ its standard deviation.

\subsection{Results}
\begin{figure}
    \centering
    \includegraphics[width=\linewidth]{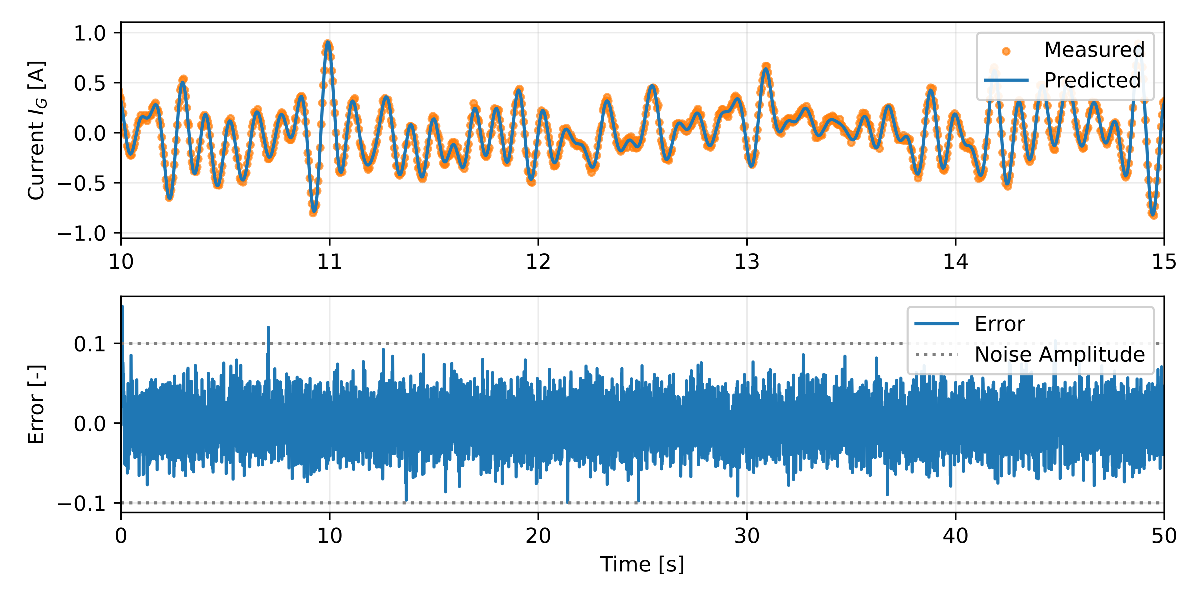}
    \caption{Simulation of the identified model. In the first subplot, the dots represent the sampled true output values and the solid line indicates the simulated model response, for five seconds for visibility. The second subplot shows the error between the predicted and measured outputs for the entire time period of 50 seconds, where the std of the noise is given by the dotted line.}
    \label{fig:pred-error-knowndist}
\end{figure}

In order to assess the capability of the proposed identification approach to capture the behaviour of the DC power network, the simulated output response of the identified model is compared with the measured data from the test set. Fig. \ref{fig:pred-error-knowndist} shows the results of the trained model on the test data set. The first subplot shows the simulated results compared to the system output containing noise. The output of the model tracks the noisy output closely. The second subplot supports this, as the error magnitudes remain within the $\pm$0.1 noise amplitude implied by an SNR of 20 dB. This confirms that despite the measurement noise, the model reproduces the measured data, indicating a good fit with an NRMS of 0.1017.

To evaluate whether the proposed method can recover the physical parameters of the DC network, we focus on the measurement set $(I_G, V_1, V_2)$. This selection is motivated by recent results on linear DAE systmes \citep{montanari2024identifiability}, which show that accurate parameter recovery requires sufficiently informative outputs.

\begin{figure}
    \centering
    \includegraphics[width=0.9\linewidth]{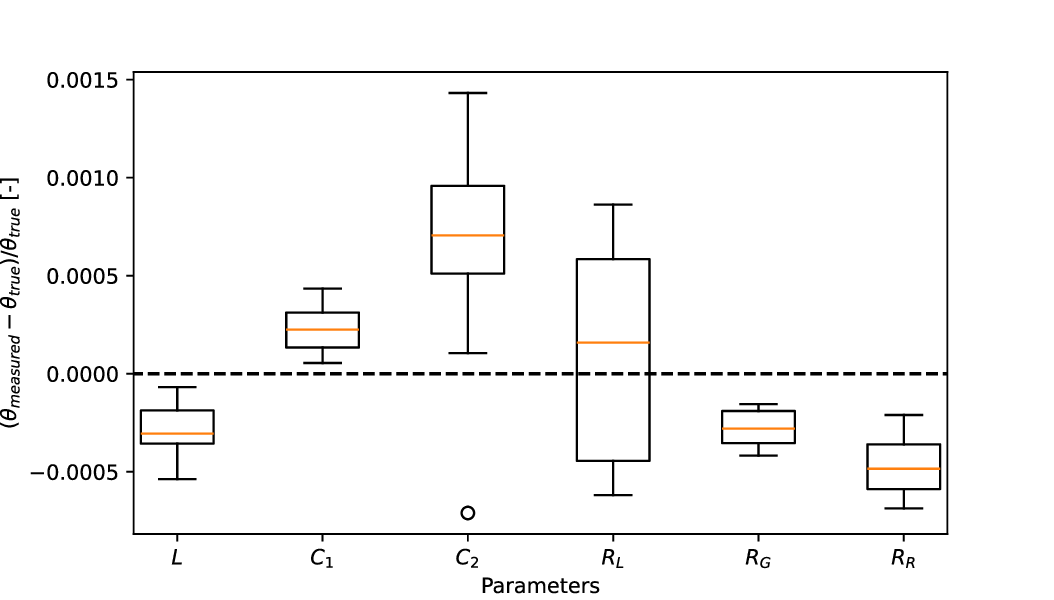}
    \caption{Boxplot of the normalised parameter deviations for 10 independent runs with noisy data sets containing measurement of $(I_G, V_1, V_2)$ with noise level of 40 dB.}
    \label{fig:boxplot}
\end{figure}

Using this output set, the identified parameters closely match their nominal values. Fig.~\ref{fig:boxplot} shows the normalised parameter deviations across ten independent simulations with noisy data (SNR = 40~dB). Most parameters show deviations in the range of 0.01\%–0.05\%, with only a few estimates approaching 0.1\%. These results confirm that the proposed method provides accurate and stable estimates of the physical parameters.
To assess the robustness of the approach against measurement noise, the same experiment is repeated for different noise levels. Table~\ref{tab:NRMS-SNR} summarises the resulting NRMS values for SNRs of 10, 20, and 30\,dB. As expected, the NRMS increases with decreasing SNR, yet the error remains proportional to the noise amplitude, showing that the method maintains consistent performance and does not overfit noisy data. Even at high noise levels, the identified models reproduce the system response with accuracy close to the noise limit, demonstrating the reliability of the proposed structure-preserving identification framework.

\begin{table}[h]
\centering
\caption{Model accuracy under different measurement noise levels.}
\label{tab:NRMS-SNR}
\begin{tabular}{ccc}
\hline
SNR (dB) & Noise std & NRMS \\
\hline
30 & 0.031 & 0.035 \\
20 & 0.100 & 0.102 \\
10 & 0.316 & 0.302 \\
\hline
\end{tabular}
\end{table}

\section{Conclusion}
\label{sec:conclusion}
This work presents an identification approach for linear port-Hamiltonian differential-algebraic equation (pH-DAE) systems based on noisy input–output data. The method combines the port-Hamiltonian framework with a backward Euler DAE solver and Newton’s method, enabling simultaneous solving of differential and algebraic components within the SUBNET identification approach. In the simulation study, the approach accurately recovered the physical parameters and reproduced the dynamic behaviour of the DC network, despite the presence of measurement noise. The results indicate that, within the considered setting, the proposed method provides reliable system identification and maintains consistent performance under different noise levels.

\bibliography{ifacconf}

\end{document}